\documentclass[aps,pra,twocolumn,superscriptaddress,floatfix]{revtex4-2}

\usepackage{amsmath,amssymb,graphicx,bbold}
\usepackage{dcolumn}
\usepackage{bm}
\usepackage{xcolor}
\usepackage{physics}
\usepackage{ulem}
\usepackage{dsfont}
\usepackage{subcaption}
\usepackage{mathtools}
\usepackage{booktabs}
\usepackage[colorlinks=true,urlcolor=black,citecolor=black,linkcolor=black]{hyperref}

\providecommand{\bra}[1]{\langle #1 \rvert}
\providecommand{\ket}[1]{\lvert #1 \rangle}

\begin{document}
	
\title{High-Dimensional Structured Light Tomography with a Tilted Lens}

\author{Altilano C. Barbosa}
\email{altilanob@id.uff.br}
\affiliation{Instituto de Física, Universidade Federal Fluminense, 24210-346 Niterói, RJ, Brazil}
\affiliation{Centro de Física de Materiales, Paseo Manuel de Lardizabal 5, 20018 Donostia-San Sebastián, Spain}

\author{Isaac Tribaldo}
\affiliation{Centro de Física de Materiales, Paseo Manuel de Lardizabal 5, 20018 Donostia-San Sebastián, Spain}

\author{Amanda K. Fritsch}
\affiliation{Instituto de Física, Universidade Federal Fluminense, 24210-346 Niterói, RJ, Brazil}

\author{Marcos G. de Oliveira}
\affiliation{Instituto de Física, Universidade Federal Fluminense, 24210-346 Niterói, RJ, Brazil}

\author{Gabriel Molina-Terriza}
\affiliation{Centro de Física de Materiales, Paseo Manuel de Lardizabal 5, 20018 Donostia-San Sebastián, Spain}
\affiliation{Donostia International Physics Center, Paseo Manuel de Lardizabal 4, 20018 Donostia-San Sebastián, Spain}
\affiliation{IKERBASQUE, Basque Foundation for Science, María Díaz de Haro 3, 48013 Bilbao, Spain}

\author{Antonio Z. Khoury}
\affiliation{Instituto de Física, Universidade Federal Fluminense, 24210-346 Niterói, RJ, Brazil}
\affiliation{Centro de Física de Materiales, Paseo Manuel de Lardizabal 5, 20018 Donostia-San Sebastián, Spain}
\affiliation{Donostia International Physics Center, Paseo Manuel de Lardizabal 4, 20018 Donostia-San Sebastián, Spain}
\date{\today}

\begin{abstract}
%In this work, we present a comprehensive study of light propagation through a tilted lens, deriving its phase transmission function, effective focal lenghts and analyzing its role as an astigmatic mode converter. We establish its explicit domain of validity within Hermite-Gaussian basis. Furthermore, we demonstrate the utility of the tilted lens as an efficient tool for high-dimensional spatial mode state tomography without restriction to fixed spatial mode order.
In this work, we investigate high-dimensional spatial mode tomography using intensity measurements and 
the astigmatic transformation implemented with a tilted lens. We improve the method previously published in 
Ref. \cite{de2025informationally} to perform the spatial tomography not restricted to a fixed order, which allowed 
us to achieve good fidelity for a mode space dimension as high as $D=66\,$. 
Since the literature is not unanimous, we also present a comprehensive revision of light propagation through a tilted 
lens for an accurate description of the transformation used for tomography. We derive its effective focal lengths, and analyze its role as an astigmatic mode converter. 
Moreover, we establish its explicit domain of validity within the Hermite-Gaussian basis. 
\end{abstract}

\maketitle

%%%%%%%%%%%%%%%%%%%%%%%%%%%%%%%%%%%%%%%%%%%%%%%%%%%%%%%%%%%%%%%%%%%%%%%%%%%%%%%%
\section{Introduction}

Structured light, characterized by tailored amplitude, phase, and polarization profiles, customized optical pulses, and recently explored spatiotemporal configurations \cite{zhan2024spatiotemporal}, has emerged as a versatile asset in optical communications, classical and quantum information processing, and optical manipulation. These modes provide natural, complete orthogonal bases for high-dimensional Hilbert spaces, enabling high-capacity quantum key distribution and spatial-mode multiplexing \cite{al2021structured}. Although a plethora of structured light families exists \cite{forbes2021structured}, Hermite--Gaussian (HG) and Laguerre--Gaussian (LG) modes form the canonical bases for optical systems possessing cartesian or cylindrical symmetries, respectively. 

In many applications, it is necessary to convert LG modes into HG modes or vice versa. This task was initially realized using pairs of cylindrical lenses \cite{cilindrical, cilindricalperformance}, a method primarily employed to measure the magnitude and sign of the orbital angular momentum (OAM) carried by an LG beam, a feature that cannot be determined from a single intensity measurement with a conventional camera. Subsequently, a simple tilted lens was introduced as a compact mode converter for OAM measurements \cite{vaity2013measuring}. Also, given a pump beam, the astigmatism from the tilted lens was recently shown to be inherited by beams generated from nonlinear effects \cite{fritsch_2026_tiltedlens}. 

The fundamental difference between the two approaches lies in structural stability: cylindrical lenses can convert modes into propagation-invariant forms, whereas full mode conversion with a tilted lens occurs only at a specific output plane. At intermediate planes, the propagating fields correspond to Generalized Hermite--Gaussian Beams (GHGBs) \cite{abramochkin2004generalized}. However, existing literature \cite{Massey:69, vaity2013measuring, buono2022eigenmodes} presents conflicting expressions for the effective focal lengths $(f_x, f_y)$ of a tilted lens, and by consequence they also disagree about the conversion planes; in this work, we resolve this discrepancy analytically, recovering the work of Massey and Siegman, and we validate the corrected focal lengths experimentally.

Building upon the ability to break OAM phase sign symmetry, the tilted lens has recently been adopted to extract relative phase information from optical fields, giving rise to intensity-only spatial state tomography \cite{braia2019, braia2021}. In these protocols, the tilted lens was modeled as a standard astigmatic mode converter. However, conventional mode converters act strictly within subspaces of a fixed total mode order $N = n + m = |l| + 2p$, rendering them incapable of resolving relative phases between modes belonging to different total-order subspaces via a single unitary transformation. Here $n,m$ and $l,p$ are the indexes of the HG modes and LG modes, respectively.  

In most applications, both classical and quantum, one needs to characterize accurately the coherence properties of the structured light source used. The standard technique involves using a 
spatial light modulator (SLM) to transform a given mode into a Gaussian profile and coupling it to a single mode optical fiber \cite{mair2001entanglement}. This method is of limited efficiency due to losses in the 
SLM modulation combined with fiber coupling. \cite{qassim2014limitations} Moreover, it does not provide information about the relative phases between different mode components.
In this paper, we demonstrate that a simple tilted lens acts as a generalized astigmatic transformation that introduces order-dependent phase shifts across different order subspaces. We derive the complete effective focal lengths and unitary operator describing the action of a tilted lens on the HG basis. Furthermore, we demonstrate theoretically and experimentally that this transformation satisfies pure-state informational completeness (PSIC), enabling full reconstruction of high-dimensional spatial states spanning multiple order subspaces without requiring complex interferometric setups.
We improve the method previously published in Ref. \cite{de2025informationally} to perform the spatial tomography not restricted to a fixed order, which allowed 
us to achieve good fidelity for a space spanned by all modes with order up to $N_{max}=10\,$, which corresponds to a space dimension as high as $D=66\,$. 

\section{The Tilted Lens}

\subsection{Light Propagation Through Tilted Surfaces}

We begin by considering the propagation of light through tilted optical surfaces. This problem was studied by G.~A.~Massey and A.~E.~Siegman in Ref.~\cite{Massey:69}, where they analyzed the reflection and refraction of Gaussian beams at tilted ellipsoidal surfaces.

Consider a biconvex lens modeled as two spherical caps with radius $R$ and cap height $d$, where $d$ corresponds to half the lens thickness. We introduce a coordinate system whose origin lies at the center of the lens. In this coordinate system, the center of each sphere is displaced by a distance $R-d$ along a direction that forms an angle $\theta$ with the optical axis, this is the incidence angle. This geometry allows us to derive relations between the coordinates $x$, $y$, and $z$ that describe the tilted spherical surfaces.
\begin{figure}[ht]
\centering
\includegraphics[width=0.8\linewidth]{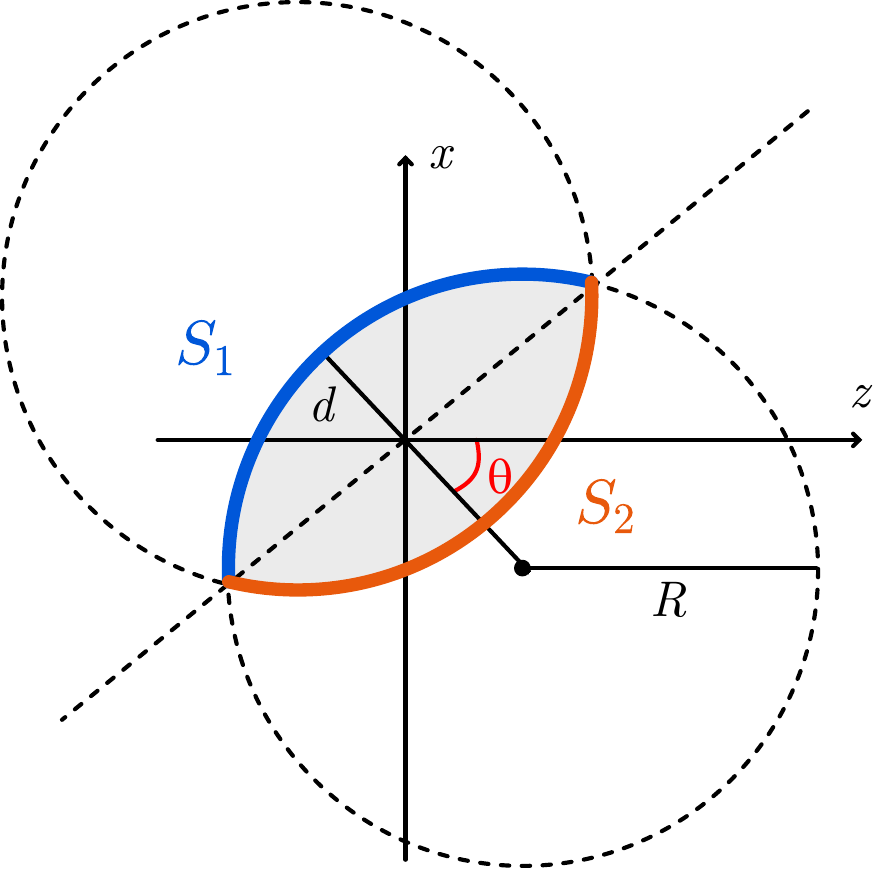}
\caption{Cross section of a symmetric biconvex lens modeled by two spherical caps of equal radius.}
\label{fig:2spherecap}
\end{figure}
Figure~\ref{fig:2spherecap} shows a cross section of the sphere in the $xz$ plane, together with the spherical cap corresponding to the lens. A laser beam propagating along the $z$ axis traverses the lens at an angle $\theta$. The phase acquired by the electric field at each point is determined by the optical path length through the lens. The ABCD matrix describing propagation through the lens entrance surface $S_1$ is given by
\begin{equation}
    \mathcal{M}_1 = diag( \mathcal{M}_T,  \mathcal{M}_S),
\end{equation}
where $\mathcal{M}_T$ and $\mathcal{M}_S$ are the ABCD matrices for the tangential and sagittal planes, respectively, with ordering $(x,\alpha_x, y, \alpha_y)$. Their elements are listed in Table~\ref{tab:ABCD}.
\begin{table}[htbp]
    \centering
    \caption{ABCD matrix elements for a tilted spherical surface\cite{Massey:69}.}
    \label{tab:ABCD}
    \begin{tabular}{|c|c|}
        \hline
        Matrix elements & Values \\
        \hline
        $A$ (tangential) & $\frac{\cos{\theta^\prime}}{\cos{\theta}}$ \\
        \hline
        $A$ (sagittal) & $1$ \\
        \hline
        $B$ (both planes) & $0$ \\
        \hline
        $C$ (tangential) & $\frac{\cos{\theta} - n_r\cos{\theta^\prime}}{Rn_r\cos{\theta}\cos{\theta^\prime}}$ \\
        \hline
        $C$ (sagittal) & $\frac{\cos{\theta} - n_r\cos{\theta^\prime}}{Rn_r}$ \\
        \hline
        $D$ (tangential) & $\frac{\cos{\theta}}{n_r\cos{\theta^\prime}}$ \\
        \hline
        $D$ (sagittal) & $\frac{1}{n_r}$ \\
        \hline
    \end{tabular}
\end{table}
Here, $\theta^\prime$ is the refracted angle determined by Snell's law $(n_2\sin{\theta^\prime} = n_1\sin{\theta})$, and $n_r = \frac{n_2}{n_1}$ is the ratio of the refractive indices of the two media, with $n_1$ denoting the refractive index outside the lens and $n_2$ that inside it.

A similar matrix, $\mathcal{M}_2$, can be derived for the interaction with the second surface $S_2$ by making the substitutions $n_r \to 1/n_r$, $R \to -R$, and interchanging $\theta$ and $\theta^\prime$. The ABCD matrix for the entire symmetric biconvex tilted lens is then given by
\begin{equation}
    \mathcal{M}_{TL} = \mathcal{M}_2 \mathcal{M}_1.
\end{equation}
Where we assumed the thin lens approximation.

We focus here on the $C$ coefficients of the resulting matrix:
\begin{subequations}
    \begin{equation}
        C_{\mathrm{tangential}} =
        \frac{-2(n_r\cos{\theta^\prime} - \cos{\theta})}{R\cos^2{\theta}}.
    \end{equation}
    \begin{equation}
        C_{\mathrm{sagittal}} =
        \frac{-2(n_r\cos{\theta^\prime} - \cos{\theta})}{R},
    \end{equation}
\end{subequations}

The effective focal lengths $f_x$ and $f_y$ can be obtained from these coefficients using $f_{x,y}=-1/C_{x,y}$, identifying the tangential and sagittal planes with the $x$ and $y$ directions, respectively. However, the resulting expressions depend on the lens type. For a plano-convex lens, the $C$ coefficients are half those of a biconvex lens. Conversely, the nominal focal length of a plano-convex lens is twice that of a biconvex lens with the same radius of curvature and refractive index. By combining these relations with the lensmaker's formula, we obtain expressions for the effective focal lengths that are valid for plano-convex and biconvex lens:
\begin{subequations}
    \begin{equation}
        f_x = f_0 \frac{(n_r - 1) \cos^2\theta}{n_r\cos\theta^\prime - \cos\theta},
    \end{equation}
    \begin{equation}
        f_y = f_0 \frac{(n_r - 1)}{n_r\cos\theta^\prime - \cos\theta}.
    \end{equation}
    \label{eqn:fxfy}
\end{subequations}
Here, $f_0$ is the nominal focal length of the lens.

For a comprehensive derivation of the ABCD ray matrices for tilted optical surfaces, readers are referred to the work of Massey and Siegman~\cite{Massey:69} and/or the excellent book Laser Resonators and Beam Propagation by Hodgson and Weber \cite{HodgsonWeber2005LaserResonators}.

%%%%%%%%%%%%%%%%%%%%%%%%%%%%%%%%%%%%%%%%%%%%%%%%%%%%%%%%%%%%%%%%%%%%%%%%%%%%%%%%
\subsection{Astigmatic Lens Operator in the Hermite--Gaussian Basis}

Here we summarize the operator derived in Appendix~\ref{appendix:derivation}, corresponding to the tilted lens followed by free propagation.
It will be useful to write it in a factorized form, which allows the independent treatment of the tilted lens action in each one of its 
principal axes and to explicitly account for the resulting independent Gouy phase contributions. 
Consider a Hermite--Gaussian ($\text{HG}_{nm}$) mode, with order $N = n+m$, evaluated at a plane located at a distance $z_0$ from its waist.
We may write it as
\begin{equation}
\begin{split}
\text{HG}_{nm}(x,y,z_0)
&= X_n(x,z_0)\,Y_m(y,z_0)\,
\\
&\times e^{-i(n+m+1)\arctan\left(\frac{z_0}{z_r}\right)}\;,
\end{split}
\end{equation}
where
\begin{equation}
\begin{split}
X_{n}(x,z_0)
&=
\frac{\mathcal{C}_{n}}{\sqrt{w(z_0)}}\,
H_n\left(\frac{\sqrt{2}x}{w(z_0)}\right)
e^{-\frac{x^2}{w^2(z_0)}}
e^{ik\frac{x^2}{2R(z_0)}}\;,
\end{split}
\end{equation}
$z_r=kw^2(0)/2$ is the Rayleigh distance, $R(z_0)=z_0+z_r^2/z_0$ is the wavefront radius and $\mathcal{C}_n$ is a normalization constant. 
An equivalent expression applies to $Y_m(y,z_0)\,$. 
The action of an astigmatic lens (such as a tilted lens) followed by free-space propagation over a distance $z$ can be described via Fresnel diffraction. Evaluating the diffraction integral (see Appendix~\ref{appendix:derivation} for full details) reveals that the transmitted field retains its Hermite--Gaussian character up to spatial coordinate rescaling and an anisotropic Gouy phase accumulation. In mathematical terms,
\begin{eqnarray}
X_n(x,z_0)\to X_n(x,z)\,e^{in\phi_x}\,,
\nonumber\\
Y_m(y,z_0)\to Y_m(y,z)\,e^{im\phi_y}\,.
\end{eqnarray}
Defining $\bar{z}_r = kw^2(z_0)/2$ and $\Delta_x = \bar{z}_r \left( \frac{1}{z} - \frac{1}{f_x} + \frac{1}{R(z_0)} \right)$, the explicit phase shift accumulated along the principal axis $x$ is given by
\begin{eqnarray}
\phi_x &=& \arctan\left( \Delta_x \right)\, -\frac{\pi}{2} - \arctan\left(\frac{z_0}{z_r}\right),
\label{eqn:phi_x}
\end{eqnarray}
with an analogous definition for $\phi_y$ upon replacing $f_x \rightarrow f_y\,$. 
Consequently, the total non-global phase factor acquired by the mode $\text{HG}_{nm}$ is
\begin{equation}
e^{i(\phi_x n + \phi_y m)} = e^{i(n-m)\frac{\phi_x-\phi_y}{2}} e^{i(n+m) \frac{\phi_x+\phi_y}{2}}.
\label{eqn:decomposed_phase}
\end{equation}
Thus, an astigmatic lens introduces two independent phase terms: one proportional to $(n-m)$, driving traditional astigmatic mode conversion, and another proportional to the total mode order $(n+m)$. The phase parameters $\phi_x$ and $\phi_y$ correspond to Gouy phases accumulated along the principal astigmatic axes.

In the HG basis, the overall transformation is governed by the unitary operator
\begin{equation}
U_{\phi_x, \phi_y} = \sum_{k} e^{i(\phi_x n_k + \phi_y m_k)} \ket{u_{k}}\bra{u_{k}},
\label{eqn:operator}
\end{equation}
where $\{u_k\}$ denotes the HG basis states and $k$ indexes the mode pairs $(n,m)$. Within a restricted space of fixed total order ($n+m = N$), the second exponential in Eq.~\eqref{eqn:decomposed_phase} reduces to an unobservable global phase, and Eq.~\eqref{eqn:operator} collapses to the familiar mode converter operator\cite{ABRAMOCHKIN1991123}:
\begin{equation}
U_N = \sum_{k} e^{i(n_k-m_k)\left(\frac{\phi_x-\phi_y}{2}\right)} \ket{u_{k}}\bra{u_{k}}\;.
\label{eqn:MCO}
\end{equation}
However, in general multi-order spaces, the full operator $U_{\phi_x,\phi_y}$ provides inter-order phase control, enabling structured light state tomography beyond fixed-order subspaces.

\subsection{Experimental Setup and Validation}
\label{sec:exp_setup}

\begin{figure}[ht]
    \centering
    \includegraphics[width=1\linewidth]{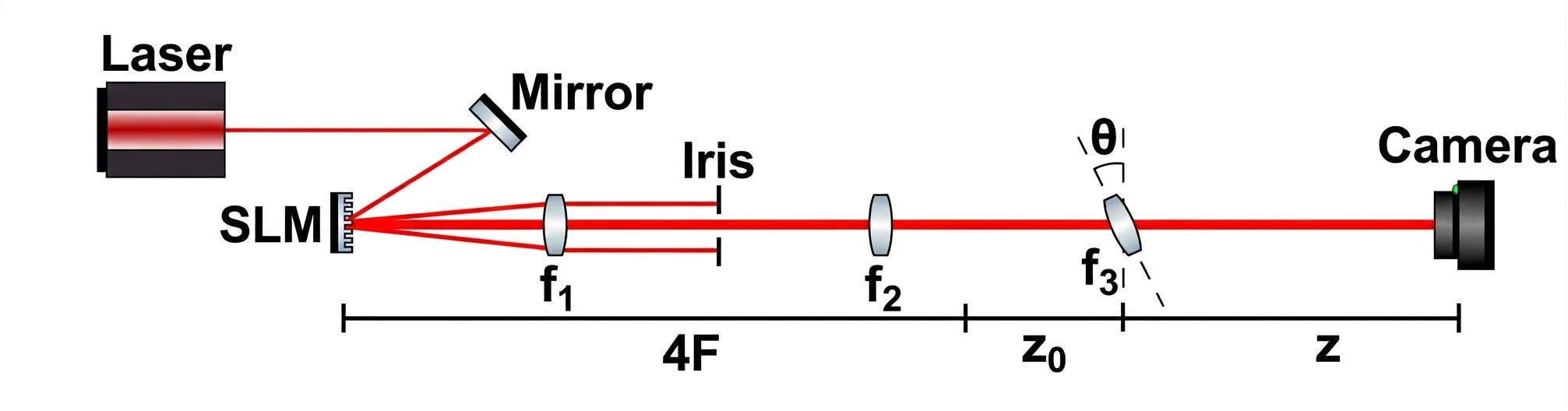}
    \caption{Experimental setup. A $633\,\mathrm{nm}$ laser beam is prepared using an SLM and a $4F$ spatial-filtering system. The field propagates over a distance $z_0 = 118\,\mathrm{cm}$ to a tilted lens and then over a distance $z$ to a CMOS camera.}
    \label{fig:setup3}
\end{figure}

To experimentally validate the transformation model described by Eq.~\eqref{eqn:operator}, we performed the experiment illustrated in Fig.~\ref{fig:setup3}. A $633\,\mathrm{nm}$ laser beam illuminates a PLUTO-2.1 LCOS spatial light modulator (SLM) from Holoeye, which generates the desired spatial modes with a beam waist of $w_0 = 0.7\,\mathrm{mm}$. The generated modes are spatially filtered and imaged using a $4F$ optical system comprising two lenses ($f_1 = f_2 = 25\,\mathrm{cm}$) and an iris placed at the focal plane. After the SLM image plane, the beam propagates over a distance $z_0 = 118\,\mathrm{cm}$ before reaching a tilted biconvex lens (LB1779-A-ML, Thorlabs; $f_3 = 30\,\mathrm{cm}$), which is oriented at a tilt angle $\theta$. The beam then propagates over a distance $z$ before its intensity profile is recorded by a Basler acA5472-17um CMOS camera.

We define the modal basis using HG modes up to a maximum order $N_{\max}$. Thus, for example, $N_{\max}=3$ includes HG modes of orders $0$, $1$, $2$, and $3$. For each $N_{\max}\in\{1,\ldots,10\}$, we generated 100 sets of random expansion coefficients $\{c_i\}$, corresponding to 100 random spatial modes. We recorded intensity images for all generated modes at different propagation distances and lens tilt angles. The propagation distances were $z\in\{0.22,0.24,0.26,0.28,0.30\}\,\mathrm{m}$, and the tilt angles were $\theta\in\{0^\circ,10^\circ,20^\circ,30^\circ\}$. For each combination of $z$ and $\theta$, we also recorded a reference Gaussian beam with the same initial waist, $w(0)=0.7\,\mathrm{mm}$.

We first investigated whether the effective focal lengths predicted by Eq.~\eqref{eqn:fxfy} agree with the experimental observations. For each tilt angle $\theta$ and propagation distance $z$, we measured the beam widths $w_x(z)$ and $w_y(z)$ using a numerical fit from the recorded Gaussian beam images and compared them with the theoretical predictions, as described in Appendix~\ref{appendix:stretch}. The results in Fig.~\ref{fig:spotsize_theovsexp} show good agreement between theory and experiment. In contrast, alternative expressions for the effective focal lengths reported in the literature, such as $f_x=f_0\cos\theta$ and $f_y=f_0/\cos\theta$, exhibited larger deviations from the experimental measurements, particularly at larger tilt angles, such as $\theta=30^\circ$.
\begin{figure*}[ht]
    \centering
    \includegraphics[width=1\linewidth]{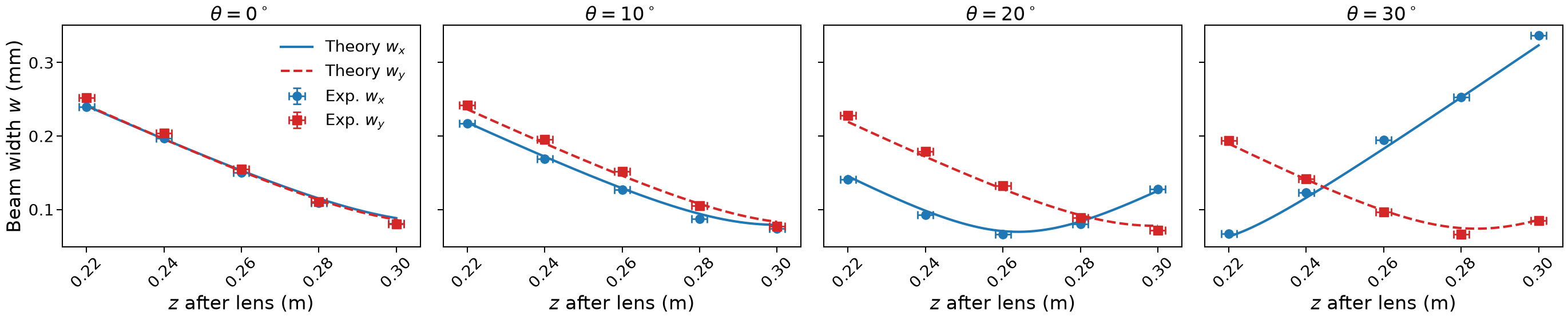}
    \caption{Comparison between theoretical and experimental beam widths for different lens tilt angles. The fitting precision is essentially determined by 2-3 times the pixel size $(2.4\mu m)$ and the corresponding error bar is not visible within graphical precision.}
    \label{fig:spotsize_theovsexp}
\end{figure*}

\begin{figure*}[ht]
    \centering
    \begin{subfigure}[b]{0.87\linewidth}
        \centering
        \includegraphics[width=\linewidth]{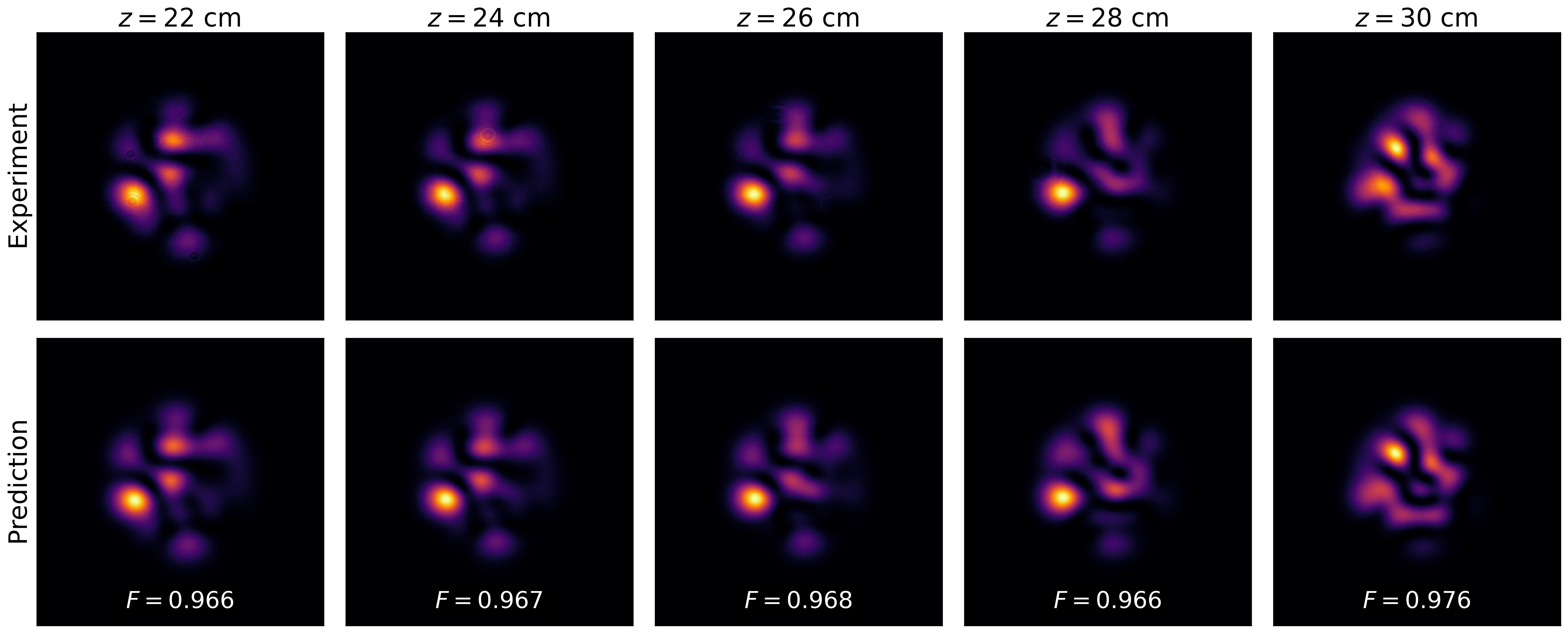}
        \subcaption{Random spatial mode with $N_{\max}=5$ and $\theta=10^\circ$.}
        \label{fig:example_random10}
    \end{subfigure}

    \vspace{1em}

    \begin{subfigure}[b]{0.87\linewidth}
        \centering
        \includegraphics[width=\linewidth]{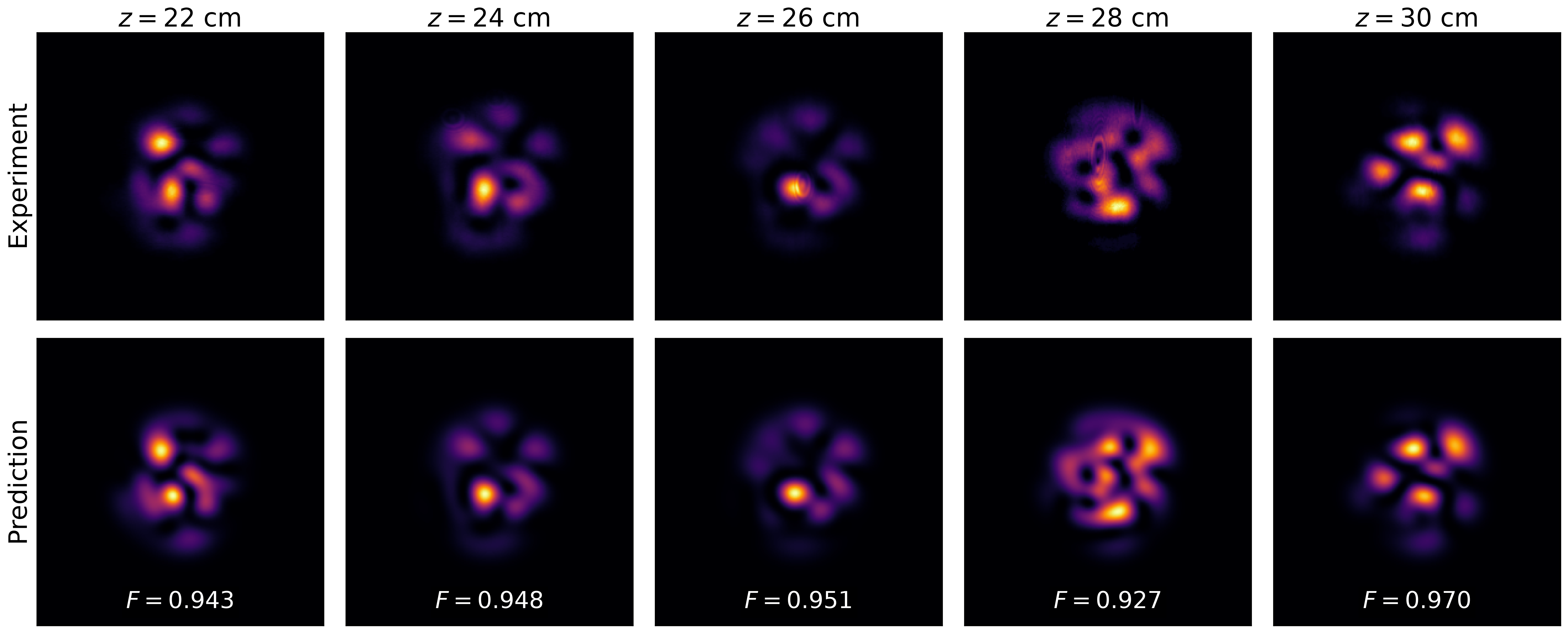}
        \subcaption{Random spatial mode with $N_{\max}=5$ and $\theta=30^\circ$.}
        \label{fig:example_random30}
    \end{subfigure}

    \caption{Comparison between numerical simulations and experimental intensity measurements for random multimode spatial fields. The propagation distance $z$ is indicated above each intensity profile. The intensity fidelity values ($\mathbb{F}$) quantify the agreement between the simulated and measured patterns.}
    \label{fig:example_random_mode}
\end{figure*}

Figure~\ref{fig:example_random_mode} compares numerical simulations based on Eq.~\eqref{eqn:operator}, using the effective focal lengths given by Eq.~\eqref{eqn:fxfy}, with experimentally measured intensity distributions. In the numerical model, we account for the transverse image inversion introduced by the 4$f$ system, which reverses both the $x$ and $y$ coordinates. Consequently, the Hermite--Gaussian modes acquire an additional mode-dependent phase factor $(-1)^{n+m}=e^{i(n+m)\pi}$. This contribution is incorporated into the corresponding $(\phi_x, \phi_y)$ phases, ensuring consistency between the analytical mode convention and the experimentally recorded coordinate system. The figure shows an example of a complex spatial mode formed by a random superposition of HG modes with orders ranging from $0$ to $5$. The intensity distributions were recorded at several propagation distances $z$, with the lens tilt angle set to $\theta=10^\circ$ and $\theta=30^\circ$.

To quantify the agreement between the simulated and experimental intensity distributions, we evaluated their intensity fidelity. Although the coordinate-stretching procedure described by Eq.~\eqref{eqn:stretchfactor} was applied to compensate for astigmatic image distortion, accurate spatial registration remained dependent on alignment references, such as the fundamental Gaussian beam. Even small spatial displacements of a few pixels can substantially reduce point-by-point overlap metrics, particularly for high-order modes with rapidly varying intensity distributions. We define the fidelity as
\begin{equation}
    \mathbb{F}(\Psi|\Phi)
    =
    \left|
    \int \Psi(\mathbf{r})\Phi^*(\mathbf{r})\,d\mathbf{r}
    \right|^2,
    \label{eqn:fidelity}
\end{equation}
where $\Psi$ and $\Phi$ represent spatial fields when calculating field/reconstruction fidelity, or intensity distributions when calculating intensity fidelity, in both cases total power is normalized to unity. In the latter case, the fidelity quantifies the overlap between the measured and simulated intensity patterns rather than the overlap between their complex fields.

\begin{figure}[ht]
    \centering
    \begin{subfigure}[b]{0.85\linewidth}
        \centering
        \includegraphics[width=\linewidth]{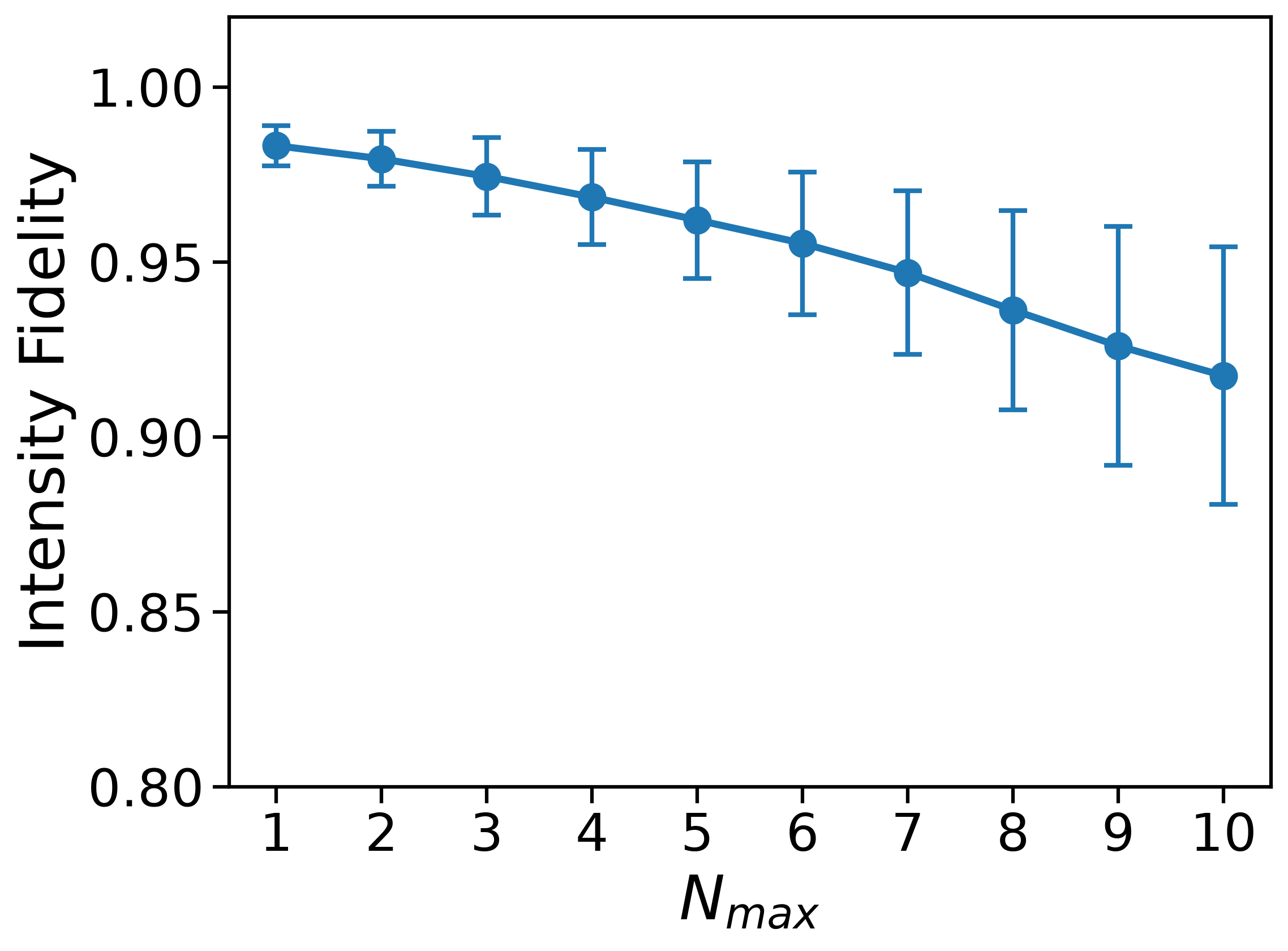}
        \subcaption{Mean intensity fidelity as a function of $N_{\max}$.}
        \label{fig:intfidN}
    \end{subfigure}

    \vspace{0.5em}

    \begin{subfigure}[b]{0.85\linewidth}
        \centering
        \includegraphics[width=\linewidth]{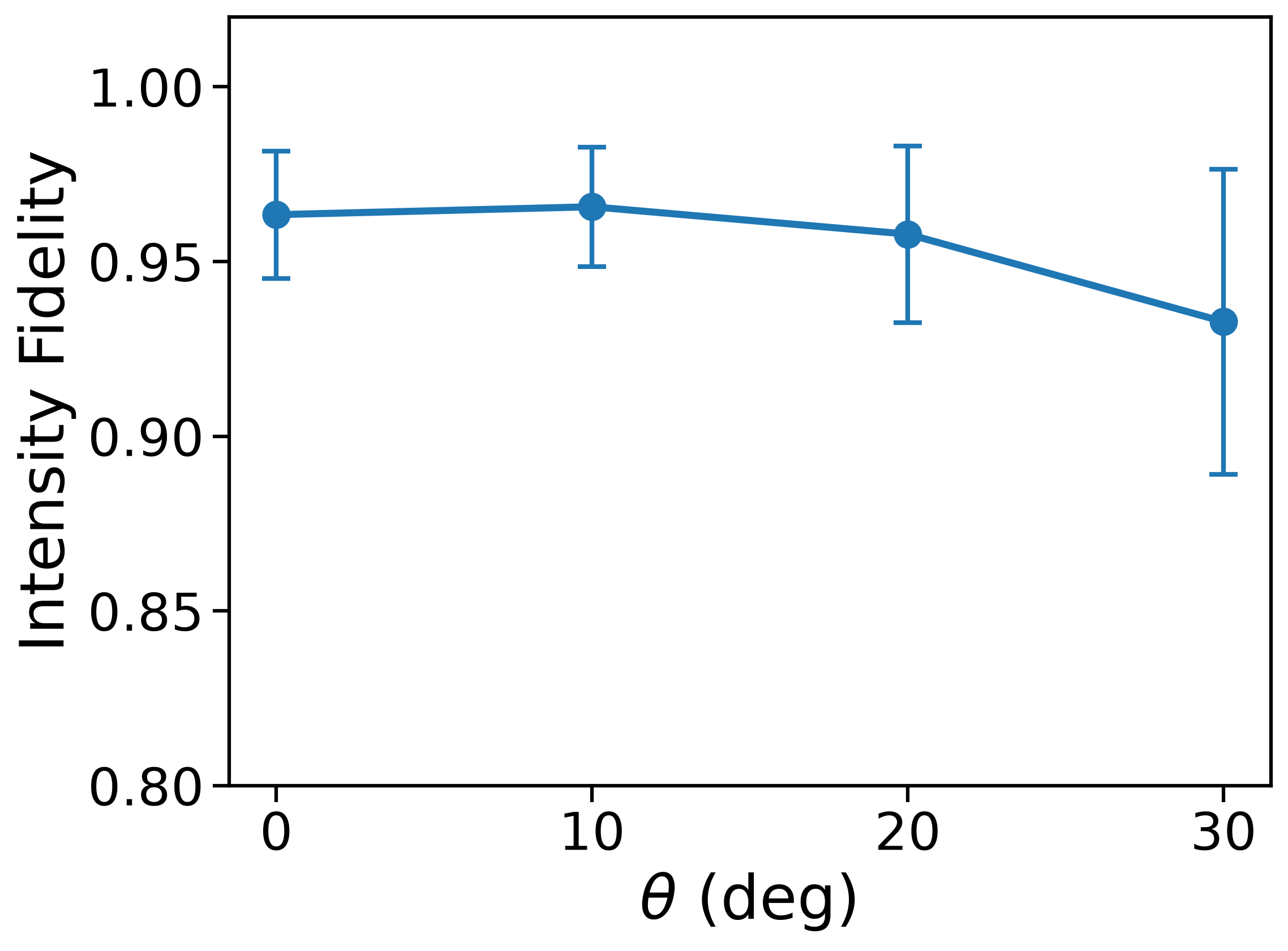}
        \subcaption{Mean intensity fidelity as a function of $\theta$.}
        \label{fig:intfidtheta}
    \end{subfigure}

    \caption{Mean intensity fidelity between the experimental measurements and the theoretical model, evaluated across all recorded modes. The results are grouped by (a) maximum HG order $N_{\max}$ and (b) lens tilt angle $\theta$. The intensity fidelity is plotted over the range [0.8, 1.0].}
    \label{fig:intfid}
\end{figure}

Figure~\ref{fig:intfid} summarizes the intensity fidelity across all recorded modes as a function of the maximum HG order $N_{\max}$ and the lens tilt angle $\theta$. The corresponding statistical results are presented in Tables~\ref{tab:intfid_N} and~\ref{tab:intfid_theta}. For each value of $N_{\max}$, the statistics comprise 2,000 comparisons, corresponding to 100 random modes measured at five propagation distances and four tilt angles. For each tilt angle, the statistics comprise 5,000 comparisons across all ten values of $N_{\max}$ and five propagation distances.

The mean intensity fidelity decreases as $N_{\max}$ increases, from $0.9832$ for $N_{\max}=1$ to $0.9174$ for $N_{\max}=10$. This trend suggests that small spatial mismatches have a greater impact on the fidelity metric for higher-order modes. Indeed, it is a result widely known in metrology \cite{santos2025optimal}. The fidelity also tends to decrease with increasing tilt angle, reaching its lowest mean value of $0.9328$ at $\theta=30^\circ$. This behavior may be attributed to the more pronounced coordinate stretching required at larger tilt angles, which can introduce additional pixel-level errors during image correction. Nevertheless, the mean intensity fidelity remains above $0.93$ for all tilt angles considered.

\begin{table}[htbp]
    \centering
    \caption{Intensity fidelity statistics as a function of the maximum HG order $N_{\max}$.}
    \label{tab:intfid_N}
    \begin{tabular}{ccccc}
        \hline
        $N_{\max}$ & Mean & Std & Minimum  \\
        \hline
        1  & 0.9832 & 0.0058 & 0.9456  \\
        2  & 0.9795 & 0.0078 & 0.9285  \\
        3  & 0.9744 & 0.0111 & 0.9058  \\
        4  & 0.9685 & 0.0136 & 0.8863  \\
        5  & 0.9619 & 0.0167 & 0.8527  \\
        6  & 0.9553 & 0.0204 & 0.8364  \\
        7  & 0.9469 & 0.0234 & 0.8066  \\
        8  & 0.9362 & 0.0285 & 0.8060  \\
        9  & 0.9260 & 0.0341 & 0.7755  \\
        10 & 0.9174 & 0.0368 & 0.7579  \\
        \hline
    \end{tabular}
\end{table}

\begin{table}[htbp]
    \centering
    \caption{Intensity fidelity statistics as a function of the lens tilt angle $\theta$.}
    \label{tab:intfid_theta}
    \begin{tabular}{ccccc}
        \hline
        $\theta$ ($^\circ$) & Mean & Std & Minimum  \\
        \hline
        0  & 0.9634 & 0.0182 & 0.8717  \\
        10 & 0.9657 & 0.0171 & 0.9002  \\
        20 & 0.9579 & 0.0253 & 0.8213  \\
        30 & 0.9328 & 0.0437 & 0.7579  \\
        \hline
    \end{tabular}
\end{table}

\section{Beam Spatial Tomography with a Tilted Lens}

In full quantum state tomography, a measurement scheme must be informationally complete (IC), enabling unique reconstruction of arbitrary mixed states of dimension $D$ described by $D \times D$ density matrices, requiring $\mathcal{O}(D^2)$ independent parameters \cite{de2025informationally}.
However, when optical fields are prepared in pure and coherent spatial states, a less restrictive condition, pure-state informational completeness (PSIC), suffices. A measurement set is PSIC if the measured outcome probabilities uniquely determine any pure state up to a global phase.
Because pure states in a $D$-dimensional Hilbert space require significantly fewer parameters to be characterized, PSIC protocols dramatically reduce the required number of measurement settings. This lowers experimental complexity while preserving full reconstruction accuracy for pure optical fields.

Our spatial tomography protocol is formulated around the PSIC criterion. Below, we lay out the mathematical framework, kernel rank criteria, and implementation in the HG basis.

\subsection{Spatial Tomography Completeness}
\label{sec:completeness}

In this section we first formulate the theoretical framework for spatial state tomography using a tilted lens. After that, we derive the measurement operators generated by the transformation and establish the algebraic conditions for PSIC.

Let $\mathcal{R}$ define a detection region in the transverse plane, in other words, a pixel from the camera. The corresponding Positive Operator-Valued Measure (POVM) element is
\begin{equation}
\Pi(\mathcal{R})=\int_\mathcal{R} |\mathbf{r}\rangle\langle\mathbf{r}|\,d\mathbf{r}\;,
\end{equation}
with matrix elements in the HG basis $\{u_j(\mathbf{r})\}$ given by
\begin{equation}
C_{jk}(\mathcal{R})
=
\langle u_j|\Pi(\mathcal{R})|u_k\rangle
=
\int_\mathcal{R} u_j^*(\mathbf{r})u_k(\mathbf{r})\,d\mathbf{r}\;.
\end{equation}
The coefficients $C_{jk}(\mathcal{R})$ quantify spatial modal overlaps inside region $\mathcal{R}$.

A tilted lens applies the unitary operation
\begin{equation}
U_{\phi_x,\phi_y}|u_k\rangle
=
e^{i(n_k\phi_x+m_k\phi_y)}
|u_k\rangle\;,
\end{equation}
which transforms the POVM element into
\begin{equation}
\Pi_{\phi_x,\phi_y}(\mathcal{R})
=
U_{\phi_x,\phi_y}^\dagger
\Pi(\mathcal{R})
U_{\phi_x,\phi_y}\;.
\end{equation}
Its matrix elements become
\begin{equation}
    [\Pi_{\phi_x,\phi_y}(\mathcal{R})]_{jk}
    =
    e^{i\Delta\phi_{jk}}
    C_{jk}(\mathcal{R})\;,
\end{equation}
where the differential phase shift is
\begin{equation}
\Delta\phi_{jk}
=
(n_k-n_j)\phi_x
+
(m_k-m_j)\phi_y\;.
\end{equation}
Thus, the tilted lens map translates index variations into configurable relative spatial phases.

The outcome probability for a state $\rho$ is given by Born's rule:
\begin{equation}
P(\mathcal{R})
=
\mathrm{Tr}
\left[
\rho\,
\Pi_{\phi_x,\phi_y}(\mathcal{R})
\right]\;.
\end{equation}
Varying tilt angles grants direct measurement access to spatial state coherences.

To evaluate completeness, consider a set of measurement operators $\{\Pi_k\}$ yielding probabilities $p_k = \mathrm{Tr}(\Pi_k\rho)$. We expand $\rho$ in the generalized Gell--Mann basis of Hermitian operators:
\begin{equation}
    \rho = \sum_i x_i B_i,
\end{equation}
where $B_i \in \{\mathds{1},\,X_{jk},\,Y_{jk},\,Z_j\}$ spans the Hermitian operators acting on the truncated $D$-dimensional Hilbert space.
The generators are defined as
\begin{align}
X_{jk}
&=
\frac{|u_j\rangle\langle u_k|
+
|u_k\rangle\langle u_j|}
{\sqrt{2}}\,,\\
Y_{jk}
&=
\frac{i\left(|u_j\rangle\langle u_k|
-
|u_k\rangle\langle u_j|\right)}
{\sqrt{2}}\,,
\end{align}
\begin{equation}
Z_j
=
\frac{1}{\sqrt{j(j+1)}} \left(
\sum_{r=1}^{j}
|u_r\rangle\langle u_r|
-
j\,|u_{j+1}\rangle\langle u_{j+1}|
\right),
\end{equation}
The off-diagonal generators ${X_{jk},Y_{jk}}$ are defined for
$1\leq j<k\leq D$, whereas the diagonal generators ${Z_j}$ are
defined for $1\leq j\leq D-1$. Together with the identity matrix,
these generators form an orthogonal basis for the space of
$D\times D$ Hermitian matrices. Their expectation values with the transformed POVM elements are
\begin{equation}
    \begin{split}
    \mathrm{Tr}\left[
    \Pi_{\phi_x,\phi_y}(\mathcal{R})X_{jk}
    \right] 
    &=
    \sqrt{2}\,
    \mathrm{Re}
    \left(
    e^{i\Delta\phi_{jk}}C_{jk}
    \right),\\
    \mathrm{Tr}\left[
    \Pi_{\phi_x,\phi_y}(\mathcal{R})Y_{jk}
    \right]
    &=
    \sqrt{2}\,
    \mathrm{Im}
    \left(
    e^{i\Delta\phi_{jk}}C_{jk}
    \right),\\
    \mathrm{Tr}\left[
    \Pi_{\phi_x,\phi_y}(\mathcal{R})Z_j
    \right]
    &= \frac{1}{\sqrt{j(j+1)}} \\
    &\times\left[\sum_{r=1}^j C_{rr} - jC_{j+1, j+1} \right].
    \end{split}
    \label{eqn:xyz}
\end{equation}

Writing $C_{jk} = |C_{jk}|e^{i\alpha_{jk}}$, we obtain for the $X_{jk}$ and $Y_{jk}$ matrices
\begin{align}
\mathrm{Tr}\left[
\Pi_{\phi_x,\phi_y}(\mathcal{R})X_{jk}
\right]
&=
\sqrt{2}|C_{jk}|
\cos(\Delta\phi_{jk}+\alpha_{jk})\,,\\
\mathrm{Tr}\left[
\Pi_{\phi_x,\phi_y}(\mathcal{R})Y_{jk}
\right]
&=
\sqrt{2}|C_{jk}|
\sin(\Delta\phi_{jk}+\alpha_{jk})\,.
\label{eqn:gell}
\end{align}
Modulating $(\phi_x,\phi_y)$ sinusoidally drives modal coherences, enabling direct extraction of both amplitude and phase coefficients. Now suppose we take two pictures with different $\phi_{x,y}\,$, one to extract amplitude and other to extract phase information. The total POVM is expanded to $\{\Pi_{\phi_x, \phi_y}, \Pi_{\phi_x^\prime, \phi_y^\prime}\}\,$, and the whole collection of traces \eqref{eqn:xyz} for both images can be stacked into a measurement matrix T.

As shown in \cite{de2025informationally}, if the transformation matrix $T$ constructed from rows $T_{ki} = \mathrm{Tr}(\Pi_k B_i)$ achieves a rank $D^2 - 1$, the measurement is informationally complete for arbitrary mixed states. For pure optical states, PSIC requires that $\mathrm{Tr}(\Pi_k|\psi\rangle\langle\psi|) = \mathrm{Tr}(\Pi_k|\phi\rangle\langle\phi|)$ for all $k$ implies $|\psi\rangle = e^{i\chi}|\phi\rangle$. Defining $S = |\psi\rangle\langle\psi| - |\phi\rangle\langle\phi|$, indistinguishable pure states satisfy
\begin{equation}
\mathrm{Tr}(\Pi_k S) = 0 \qquad \forall k\,.
\end{equation}
Because the difference between two rank-one projection operators has a rank of at most two, the POVM is PSIC if and only if the kernel (null space) of matrix $T$ contains no non-zero Hermitian matrix with $\text{rank} \le 2$ \cite{rank2}.

Expanding $S = \sum_i s_i B_i$, the condition simplifies to the linear system
\begin{equation}
T\mathbf{s}=0, \qquad T_{ki} = \mathrm{Tr}(\Pi_k B_i).
\label{eqn:T}
\end{equation}
and we can identify the vector $\mathbf{s} = (s_1, s_2,...,s_{D^2-1})$ as the kernel of T.

Therefore, the tilted-lens measurement setup is pure-state informationally complete if and only if every non-zero matrix in the null space of $T$ has a rank of at least three \cite{rank2, rank2.2}.

\subsection{Experiment and Results}

To demonstrate state tomography across multiple mode orders, we applied the tilted-lens protocol to reconstruct arbitrary structured light beams. From the large dataset acquired using the experimental setup shown in Fig.~\ref{fig:setup3}, we selected two sets of images corresponding to $(z=0.30\,\mathrm{m}, \theta=0^\circ)$ and $(z=0.22\,\mathrm{m}, \theta=20^\circ)$, respectively. All other experimental parameters were kept constant, and the images were post-processed using the anisotropic rescaling of Eq.\eqref{eqn:stretchfactor}, as explained in Appendix~\ref{appendix:stretch}. Finally, all images were uniformly downsampled to a resolution of $232\times232$ pixels.

The complete measurement test described in Sec.~\ref{sec:completeness} shows that these measurements do not provide full information for arbitrary mixed states when $N_{\max}=10$. In other words, the measurements are insufficient to uniquely reconstruct incoherent mixtures of modes. However, the results demonstrate that our choice of measurements with two images is informationally complete for pure states (PSIC).
We constructed the matrix $T$ according to Eq.~\eqref{eqn:T}, for two images and $N_{\max} = 10$. Numerical optimization did not identify any rank-2 or lower matrices in the null space of $T$. Therefore, all pure states within the subspace defined by $N_{\max}=10$ can be uniquely reconstructed. We note that the dimension of this subspace is
$D=(N_{\max}+1)(N_{\max}+2)/2=66$,
meaning that the state is described by 66 complex coefficients, or equivalently 132 real parameters before accounting for normalization and global-phase constraints. Thus, the dimension of the measured subspace scales as $\mathcal{O}(N_{\max}^2)$.

To reconstruct the spatial mode coefficients from the experimental images, we employed a maximum-likelihood estimation algorithm based on the unitary model given by Eq.~\eqref{eqn:operator}. The core of the algorithm is the minimization of the negative log-likelihood, defined under a Gaussian noise assumption as \cite{von2011bayesian, smolin2012efficient}
\begin{equation}
    -\log{\mathcal{L}}
    =
    \frac{1}{2}\sum_p
    \left(\frac{I^{\mathrm{data}}_p-I^{\mathrm{estimated}}_p}{\sigma_p}\right)^2,
\end{equation}
where $I^{\mathrm{data}}_p$ denotes the intensity measured by the camera in each pixel $p$, $I^{\mathrm{estimated}}_p$ is the intensity predicted by our model and $\sigma^2_p$ is the noise variance, which was taken as uniform unit ($\sigma_p = 1$). We refer to this negative log-likelihood as the residual cost function. The optimization was performed using a multi-start strategy with five random initializations, and the solution with the lowest residual cost was selected.

\begin{figure}[ht]
    \centering
    \includegraphics[width=0.9\linewidth]{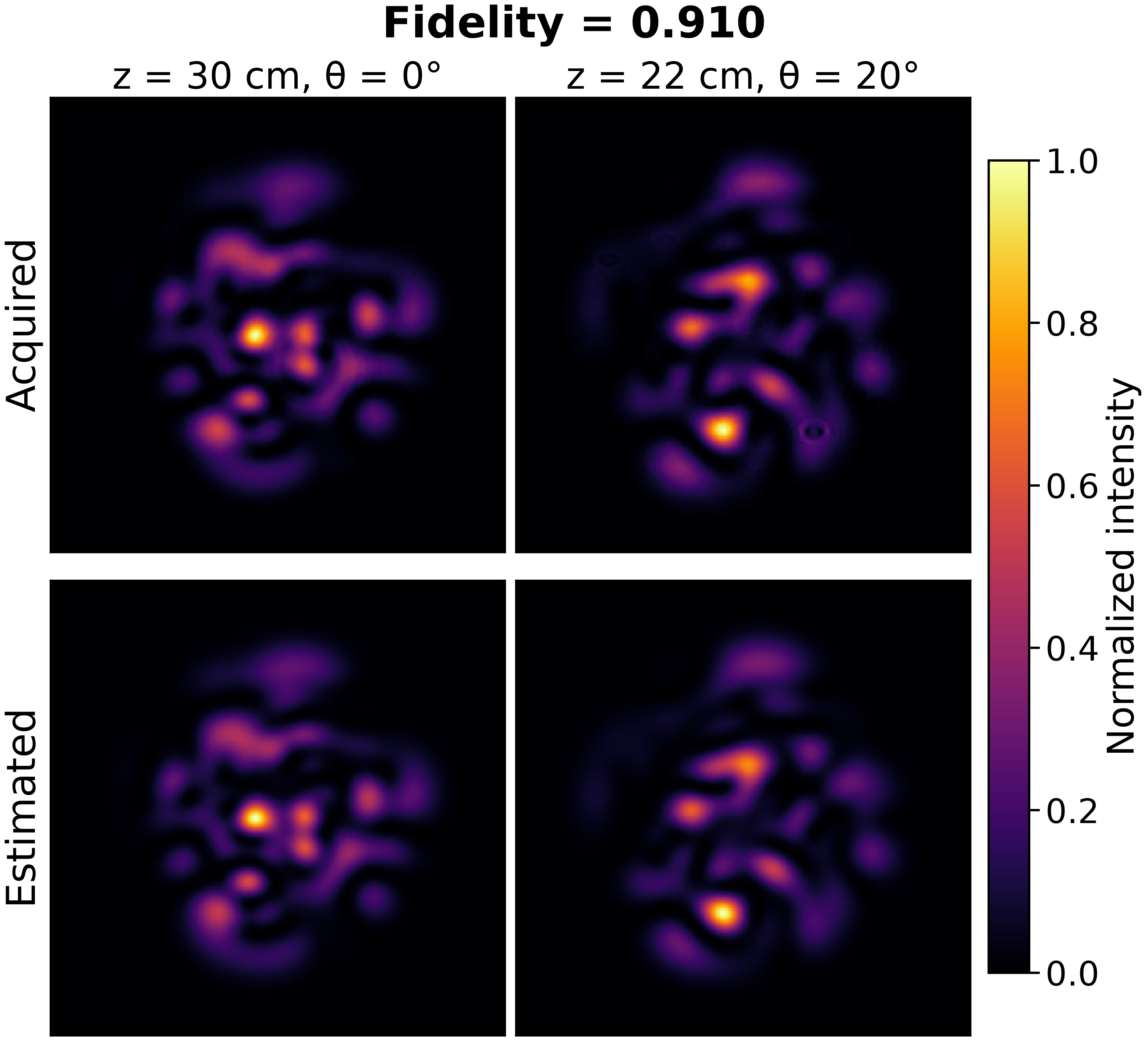}
    \caption{Comparison between the measured and reconstructed intensity profiles for a random spatial mode with $N_{\max}=10$. Here the reconstruction fidelity is equal to $91\%$, which is the median of our dataset.}
    \label{fig:estimated_K2}
\end{figure}

The reconstructed states achieved, across all our dataset, a mean reconstruction fidelity of $96.14\%$, as defined by \eqref{eqn:fidelity}, with $\Psi$ as the simulated field and $\Phi$ as the estimated field. The remaining discrepancy is attributed to residual image misalignment, uncertainties in the propagation distances $z_0$ and $z$, and imperfections in the generation of the target modes by the SLM. Figure~\ref{fig:estimated_K2} compares the measured and reconstructed intensity profiles for a random mode with $N_{\max}=10$. The reconstructed intensity distributions were calculated using Eq.~\eqref{eqn:operator} and the estimated mode coefficients.

\begin{figure}[ht]
    \centering
    \includegraphics[width=0.9\linewidth]{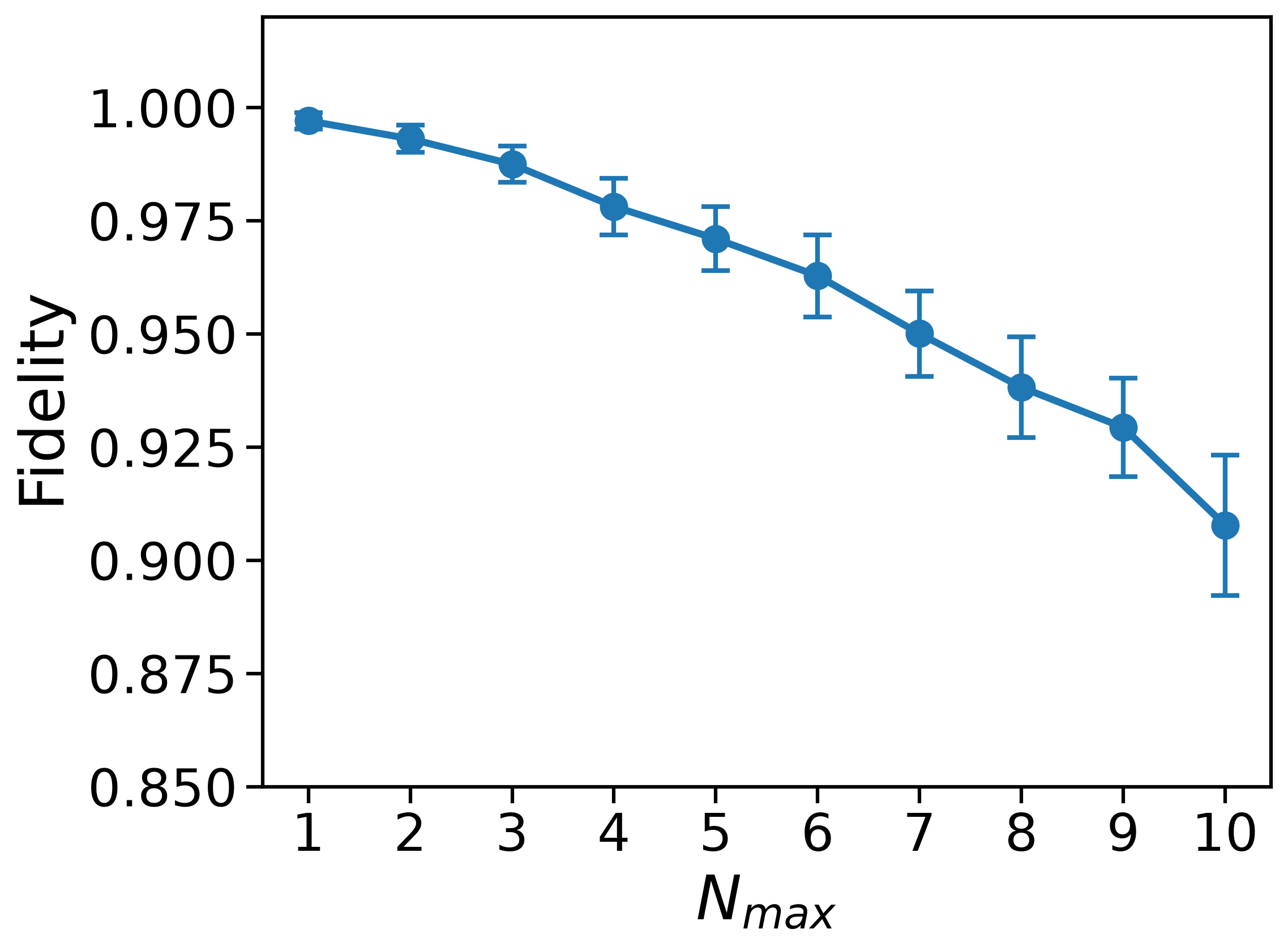}
    \caption{Mean reconstruction fidelity as a function of the maximum HG order $N_{\max}$. The reconstruction fidelity is plotted over the range [0.85, 1.0]}
    \label{fig:fid_N}
\end{figure}

Table~\ref{tab:statistics_by_N} presents the mean and standard deviation of the reconstruction fidelity and residual cost across the 100 random modes for each value of $N_{\max}$, up to $N_{\max}=10$. Figure~\ref{fig:fid_N} shows the mean fidelity as a function of $N_{\max}$.
\begin{table}[htbp]
    \centering
    \caption{Mean and standard deviation of the reconstruction fidelity and residual cost for each value of $N_{\max}$.}
    \label{tab:statistics_by_N}
    \begin{tabular}{c|cc|cc}
        \toprule
        & \multicolumn{2}{c|}{Fidelity}
        & \multicolumn{2}{c}{Cost} \\
        \cmidrule(lr){2-3} \cmidrule(lr){4-5}
        $N_{\max}$ & Mean & Std. dev. & Mean & Std. dev. \\
        \midrule
        1  & 0.9970 & 0.0017 & 2.5664  & 1.3132 \\
        2  & 0.9930 & 0.0030 & 2.6672  & 1.3286 \\
        3  & 0.9874 & 0.0040 & 3.2686  & 1.6160 \\
        4  & 0.9781 & 0.0062 & 3.3425  & 1.3872 \\
        5  & 0.9710 & 0.0070 & 4.5708  & 2.0299 \\
        6  & 0.9628 & 0.0090 & 4.6848  & 2.2827 \\
        7  & 0.9500 & 0.0094 & 5.3928 & 2.3406 \\
        8  & 0.9381 & 0.0111 & 5.9592 & 2.0728 \\
        9  & 0.9293 & 0.0108 & 6.4864 & 2.6021 \\
        10 & 0.9077 & 0.0155 & 8.0108 & 3.4289 \\
        \bottomrule
    \end{tabular}
\end{table}

Using only two intensity images yielded a mean reconstruction fidelity of approximately $91\%$ for $N_{\max}=10$. The remaining $9\%$ loss in fidelity can be attributed to numerical errors and experimental imperfections, such as misalignment, imperfect mode generation by the SLM, camera noise, and aberrations introduced by optical elements along the beam path.

To investigate the effect of incorporating additional measurements, we also considered up to six intensity images in the likelihood model. The four additional images corresponded to $(z=0.24\,\mathrm{m}, \theta=0^\circ)$, $(z=0.22\,\mathrm{m}, \theta=10^\circ)$, $(z=0.30\,\mathrm{m}, \theta=10^\circ)$, and $(z=0.30\,\mathrm{m}, \theta=20^\circ)$, which were sequentially incorporated into the estimation algorithm in the order listed. For the $N_{max}=10$ case, the additional experimental data improved the convergence of the optimization, resulting in a modest increase in the mean reconstruction fidelity, from approximately $91\%$ to $92.2\%$ when using six images, together with a slight reduction in the corresponding standard deviation.

\begin{figure}[ht]
    \centering
    \includegraphics[width=0.9\linewidth]{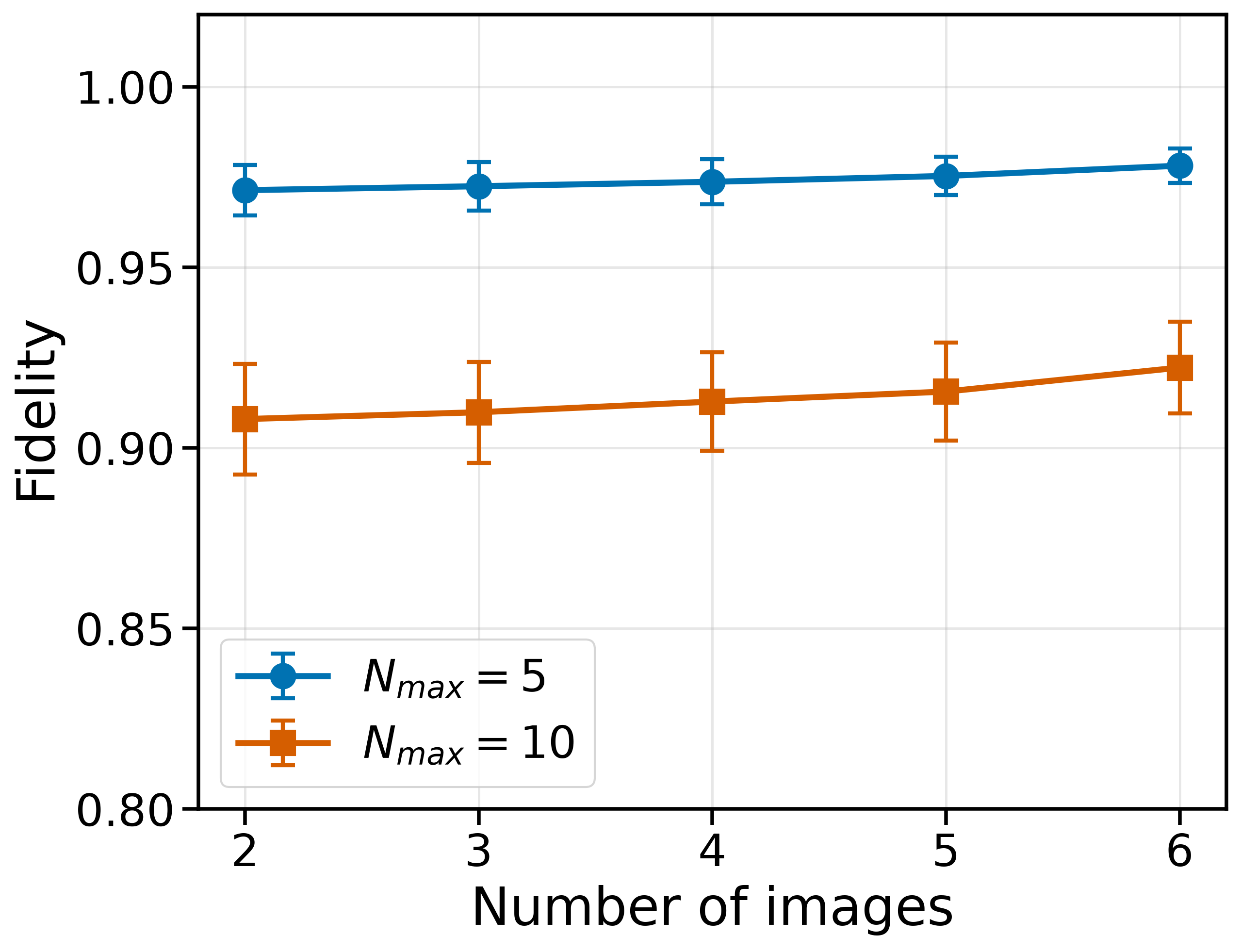}
    \caption{Mean reconstruction fidelity as a function of the number of intensity images used for coefficient estimation, for $N_{\max}=5$ and $N_{\max}=10$. The reconstruction fidelity is plotted over the range [0.8, 1.0]}
    \label{fig:fid_num}
\end{figure}

Figure~\ref{fig:fid_num} shows the mean reconstruction fidelity as a function of the number of intensity images used to estimate the mode coefficients, for $N_{\max}=5$ and $N_{\max}=10$. Both curves exhibit an increasing trend in fidelity as more images are incorporated, indicating that additional intensity measurements provide complementary information about the spatial state. However, although incorporating additional images appears to mitigate numerical errors, it does not compensate for experimental imperfections.

\section{Conclusion}
\vspace{-0.2cm}

We investigate high-dimensional structured light tomography and informational completeness with both, direct and astigmatic intensity measurements. 
The astigmatic transformation was implemented with a tilted lens. In view of the conflicting descriptions available in the literature, as well as 
their limitations to accurately describe the experimental results, we have presented a detailed analytical and experimental study of spatial mode transformations induced by a tilted lens. 
By accounting for anisotropic focal lengths and Gouy phase shifts, we derived the full unitary transformation matrix acting on the Hermite--Gaussian basis across multiple order spaces. We demonstrated that a tilted lens breaks the fixed-order constraint of standard astigmatic mode converters, providing an efficient route for high-dimensional spatial tomography. Using a pure-state informationally complete (PSIC) framework, we successfully reconstructed complex spatial modes up to Hilbert space dimensions of $D=66$ with good fidelity. This simple setup offers an accessible, robust tool for high-dimensional quantum optics and structured light characterization.

\vspace{-0.25cm}
\section*{Acknowledgements}
\vspace{-0.25cm}
	
Funding was provided by Conselho Nacional de Desenvolvimento Científico e Tecnológico (CNPq), Fundação Carlos Chagas Filho de Amparo à Pesquisa do Estado do Rio de Janeiro (FAPERJ), Fundação de Amparo à Pesquisa do Estado de São Paulo (FAPESP, grants 2021/06823-5 and 2022/15035-3).
Gabriel Molina-Terriza and Isaac Tribaldo acknowledge support from the CSIC Research Platform on Quantum Technologies, from IKUR Strategy under the collaboration agreement between Ikerbasque Foundation and DIPC/MPC on behalf of the Department of Education of the Basque.
Departamento de Educacion, Basque Government. IKUR Quantum Technologies. PTI-01 Quantum Technologies, CSIC. Proyecto PID2022-143268NB-I00 financiado por MCIN	
Spanish Centre for the Development of Industrial Technology (CDTI) through the Red Cervera Programme, Project QUIXOTE, under Grant No. CER-2025102 

\medskip
\noindent\textit{Use of AI tools.} During the preparation of this work, the authors used Claude Sonnet 5.5 to polish the text of the manuscript and as a debugging aid for the code used in the estimation procedure. The authors reviewed and edited all AI-assisted output and take full responsibility for the content of the paper.

\vspace{-0.25cm}
\section*{Author Contributions}
\vspace{-0.25cm}
Author contributions are reported using the CRediT (Contributor Roles Taxonomy) system.\\[0.3em]
\textbf{A.C.B.:} Conceptualization; Formal Analysis; Investigation; Software; Writing - original draft\\
\textbf{I.T.:} Conceptualization; Investigation; Writing – review and editing \\
\textbf{A.K.F.:} Conceptualization; Investigation; Writing – review and editing \\
\textbf{M.G.O.:} Conceptualization\\
\textbf{G.M.-T.:} Funding acquisition; Supervision; Writing – review and editing\\
\textbf{A.Z.K.:} Funding acquisition; Supervision; Writing – review and editing

\vspace{-0.25cm}
\section*{Data Availability}
\vspace{-0.25cm}

The processed data, the used Maximum Likelihood scripts and PSIC tests are available at \cite{C_Barbosa_tilted_lens_code_2026}. The Maximum Likelihood scripts have also been organized into a Python package to facilitate the general use of our method, which is available at \cite{astimator}. The simulations of tilted-lens systems based on the ABCD matrix formalism are available at \cite{fritsch_2026_tiltedlens-github}. The programs used to generate the SLM holograms are available at \cite{Barbosa_structured_optics_Package_for_2026}. The raw data are available from the corresponding author upon reasonable request.
\appendix

\vspace{-0.25cm}
\section{Derivation of the Astigmatic Lens Operator in the HG Basis}
\label{appendix:derivation}
\vspace{-0.25cm}

Here we provide the detailed derivation of the field transformation produced by an astigmatic lens acting on a Hermite--Gaussian mode. Following the astigmatic lens and free-space propagation over a distance $z$, the optical field is given by Fresnel diffraction\cite{goodman1969introduction, BornWolf}:
\begin{eqnarray}
E(x,y,z)
&=& E_0\frac{C_{n}C_{m}}{w(z_0)}\,\frac{e^{ikz}}{i\lambda z}
\\
&\times&\!\!\!\iint\!\!
e^{\frac{ik}{2z}\left[(x-\xi)^2+(y-\eta)^2\right]}
e^{-\frac{ik}{2}\left(\frac{\xi^2}{f_x}+\frac{\eta^2}{f_y}\right)}
\nonumber\\
&\times&\!\!
H_n\!\left(\frac{\sqrt{2}\xi}{w(z_0)}\right)\!
H_m\!\left(\frac{\sqrt{2}\eta}{w(z_0)}\right)\!
e^{-\frac{\xi^2+\eta^2}{w^2(z_0)}}
\nonumber\\
&\times&\!\!\!
e^{ik\frac{\xi^2+\eta^2}{2R(z_0)}}
%\\
%&\times
e^{-i(n+m+1)\arctan\left(\frac{z_0}{z_r}\right)}
\,d\xi d\eta\,,
\nonumber
\end{eqnarray}
where $C_p = \frac{\sqrt[4]{2/\pi}}{\sqrt{2^{p}p!}}\,$ is a normalization constant.

Since the variables are separable, 
\begin{equation}
E(x,y,z) = E_0\,e^{ikz}F_x(x,z)\,F_y(y,z)\,,
\end{equation}
we evaluate the $x$-dependent component:
\begin{equation}
\begin{split}
F(x,z)
&= \frac{C_{n}}{\sqrt{i\lambda\,z\,w(z_0)}}
e^{-i(n+\frac12)\arctan\left(\frac{z_0}{z_r}\right)}
e^{i\frac{kx^2}{2z}}
\\
&\times
\int
e^{-A\xi^2+B\xi}
H_n\left(\frac{\sqrt{2}\xi}{w(z_0)}\right)d\xi\;,
\end{split}
\end{equation}
where we used the following definitions
\begin{eqnarray}
A &=& \frac{1 - i\Delta_x}{w^2(z_0)}\;,
\nonumber\\
B &=& -\frac{ikx}{z}\;,
\\
\Delta_x &=& \bar{z}_r \left( \frac{1}{z} - \frac{1}{f_x} + \frac{1}{R(z_0)} \right)\;,
\nonumber
\end{eqnarray}
with
\begin{eqnarray}
\bar{z}_r &\equiv& \frac{k w^2(z_0)}{2} = \frac{z_0}{z_r}R(z_0)\,,
\nonumber\\
R(z_0) &=& z_0+\frac{z_r^2}{z_0}\;.
\end{eqnarray}

Using the generating function representation for Hermite polynomials,
\begin{equation}
H_n(x) = \frac{\partial^n}{\partial t^n} \left( e^{-t^2+2tx} \right) \Bigg|_{t=0},
\end{equation}
we interchange derivatives and integrals to obtain:
\begin{equation}
\begin{split}
F(x,z)
&= \frac{C_{n}}{\sqrt{i\lambda\,z\,w(z_0)}}
\\
&\times e^{-i(n+\frac12)\arctan\left(\frac{z_0}{z_r}\right)}
e^{i\frac{kx^2}{2z}}
e^{AC^2}
\\
&\times
\frac{\partial^n}{\partial t^n}
\left(
e^{-t^2}
\int
e^{-A(\xi-C)^2}d\xi
\right)
\Bigg|_{t=0},
\end{split}
\end{equation}
where $C = \frac{1}{2A}\left(B+\frac{2\sqrt2t}{w(z_0)}\right)$. The Gaussian integral  can be 
easily evaluated, yielding
\begin{equation}
\begin{split}
F(x,z)
&=
\frac{C_{n}}{\sqrt{i\lambda\,z\,w(z_0)}}\, \sqrt{\frac{\pi}{A}}
\\
&\times e^{-i(n+\frac12)\arctan\left(\frac{z_0}{z_r}\right)}
e^{i\frac{kx^2}{2z}}
e^{\frac{B^2}{4A}}
\\
&\times
\frac{\partial^n}{\partial t^n}
\left(
e^{-\Lambda^2t^2
+
2t\frac{B}{\sqrt2Aw(z_0)}}
\right)
\Bigg|_{t=0},
\end{split}
\end{equation}
where 
\begin{eqnarray}
\Lambda &=& \sqrt{1-\frac{2}{Aw^2(z_0)}} = i\left(\frac{1+i\Delta_x}{\sqrt{1+\Delta_x^2}}\right)
\nonumber\\
&=& e^{i\left[\frac{\pi}{2}+\arctan(\Delta_x)\right]}\,.
\end{eqnarray}
Once again, we recognize the generating function to obtain
\begin{eqnarray}
    F(x,z) &=& \frac{C_{n}}{\sqrt{i\lambda\,z\,w(z_0)}}\,\sqrt{\frac{\pi}{A}}\, H_n\!\left( \frac{B}{\sqrt2\,w(z_0)A\Lambda} \right)
    \nonumber\\
    &\times& \Lambda^n     e^{\frac{B^2}{4A}}\, e^{i\frac{kx^2}{2z}}
    e^{-i(n+\frac12)\arctan\left(\frac{z_0}{z_r}\right)}\;.
    \label{eqn:finalX}
\end{eqnarray}
By isolating the index-dependent phase contribution, we find
\begin{eqnarray}
    F(x,z) &=& X_n(x,z)\,e^{in\phi_x}\;,
    \label{eqn:finalX2}
\end{eqnarray}
with
\begin{eqnarray}
X_n(x,z) &=& \frac{C_{n}}{\sqrt{i\lambda\,z\,w(z_0)}}\,\sqrt{\frac{\pi}{A}}\, H_n\!\left(\frac{kxw(z_0)}{z\sqrt{2}\sqrt{1+\Delta_x^2}}  \right)
\nonumber\\
&\times& e^{\frac{B^2}{4A}}\, e^{i\frac{kx^2}{2z}}e^{-\frac{i}{2}\arctan\left(\frac{z_0}{z_r}\right)}\;,
\label{eqn:indexindependent}
\end{eqnarray}
and
\begin{equation}
    \phi_x = \arctan\left( \Delta_x \right)\, -\frac{\pi}{2} - \arctan\left(\frac{z_0}{z_r}\right),
\end{equation}
which leads directly to the $x$ axis Gouy phase shift $\phi_x$ defined in Eq.~\eqref{eqn:phi_x}.
The derivation of the $y$ axis Gouy phase shift $\phi_y$ is completely analogous.

\section{Coordinate Stretching Correction}
\label{appendix:stretch}

In this appendix, we analyze the spatial scaling induced by the astigmatic lens and demonstrate how to numerically un-stretch the recorded intensity profiles.

First, we examine the argument of the Hermite polynomial in Eq.~\eqref{eqn:indexindependent}:
\begin{equation}
	\xi_x = \frac{k w(z_0) x}{z \sqrt{2} \sqrt{1 + \Delta_x^2}}
          = \frac{\sqrt2x}{w_x(z)}\,,
	\label{eqn:hermite_arg_simplified}
\end{equation}
where 
\begin{equation}
w_x(z) = w(z_0) \frac{z}{\bar{z}_r} \sqrt{1 + \Delta_x^2}\;,
\label{eqn:stretch}
\end{equation}
is the beam width along the $x$ direction. 
Note from Eq.~\eqref{eqn:hermite_arg_simplified} that $\xi_x$ is strictly real-valued. Therefore, the astigmatic transformation rescales the transverse coordinate $x$ without introducing complex arguments into the Hermite polynomials, preserving the structural functional form of the original modes.

Next, consider the quadratic exponential term $\exp(B^2 / 4A)$ in Eq.~\eqref{eqn:indexindependent}:
\begin{equation}
    \begin{split}
    	\frac{B^2}{4A} &= -\frac{k^2 x^2}{4 z^2 \frac{1}{w^2(z_0)}(1 - i\Delta_x)} \\
        &= -\frac{1}{w^2(z_0)} \frac{\bar{z}_r^2}{z^2} \left( \frac{1 + i\Delta_x}{1 + \Delta_x^2} \right) x^2,
    \end{split}
    \label{eqn:stretch}
\end{equation}
where we used $\bar{z}_r = k w^2(z_0)/2\,$. 

The imaginary part of Eq.~\eqref{eqn:stretch} represents the quadratic phase associated with wavefront curvature along the $x$-axis. 
This becomes clear if write
\begin{eqnarray}
	\text{Im}\left( \frac{B^2}{4A} \right) + \frac{kx^2}{2z} &=&
    \frac{k\,x^2}{2z}\left[1-\left(\frac{\Delta_x}{1+\Delta_x^2}\right)\frac{\bar{z}_r}{z}\right]
    \nonumber\\
&=& \frac{k\,x^2}{2R_x(z)}\,,
\end{eqnarray}
where
\begin{eqnarray}
	R_x(z) &=& z\left[1-\left(\frac{\Delta_x}{1+\Delta_x^2}\right)\frac{\bar{z}_r}{z}\right]^{-1}
    \nonumber\\
    &=& \frac{z^2(1+\Delta_x^2)}{(1+\Delta_x^2)z-\Delta_x\bar{z}_r}\;.
\end{eqnarray}
The real part, 
$\text{Re}(B^2 / 4A)$, determines the Gaussian envelope intensity profile along $x\,$,
\begin{equation}
	\text{Re}\left( \frac{B^2}{4A} \right) = -\frac{x^2}{w_x^2(z)}\;.
\end{equation}
Note that this effective beam width is consistent with the one appearing in Eq.\eqref{eqn:hermite_arg_simplified}. 
Therefore, the mode function after transmission through the tilted lens and free propagation becomes 
\begin{eqnarray}
X_n(x,z) &=& \frac{C_{n}}{\sqrt{i\lambda\,z\,w(z_0)}}\,\sqrt{\frac{\pi}{A}}\, H_n\!\left( \frac{\sqrt2x}{w_x(z)} \right)
\nonumber\\
&\times& e^{-\frac{x^2}{w_x^2(z)}}e^{i\frac{kx^2}{2R_x(z)}}e^{-\frac{i}{2}\arctan\left(\frac{z_0}{z_r}\right)}\;.
\end{eqnarray}
Finally, we work out the multiplicative factor
\begin{equation}
    \sqrt{\frac{\pi/A}{i\lambda\,z\,w(z_0)}} = \sqrt{\frac{\pi w^2(z_0)/\lambda}{iw(z_0)z(1-i\Delta_x)}}
%\nonumber\\ &=& 
    =\frac{e^{\frac{i}{2}\left(\frac{\pi}{2}+\arctan\Delta_x\right)}}{i\sqrt{w_x(z)}}\,,
\end{equation}
which leads to 
\begin{equation}
X_n(x,z) = \frac{C_{n}\,e^{\frac{i}{2}\phi_x}}{i\sqrt{w_x(z)}}\, H_n\!\left( \frac{\sqrt2x}{w_x(z)} \right)e^{-\frac{x^2}{w_x^2(z)}}
e^{i\frac{kx^2}{2R_x(z)}}\;,
\end{equation}
showing that, apart from mode-independent global phase shifts, the mode function retains its functional form with transformed 
beam width and wavefront radius.

An analogous derivation along the $y$-axis yields the transverse waist spot size,
\begin{equation}
	w_y(z) = w(z_0) \frac{z}{\bar{z}_r} \sqrt{1 + \Delta_y^2}\;,
    \label{eqn:unstretch}
\end{equation}
with $\Delta_y = \bar{z}_r \left( \frac{1}{z} - \frac{1}{f_y} + \frac{1}{R(z_0)} \right)\,$. 

Because $f_x \neq f_y$, the asymmetry between $w_x(z)$ and $w_y(z)$ induces an anisotropic spatial stretching factor,
\begin{equation}
	S_{x/y} = \frac{w_x(z)}{w_y(z)} = \sqrt{\frac{1 + \Delta_x^2}{1 + \Delta_y^2}}\;.
    \label{eqn:stretchfactor}
\end{equation}
To perform state tomography, the recorded experimental intensity patterns are numerically rescaled along the axis with the larger beam waist, using a scaling factor of $S_{x/y}^{-1}$ or $S_{x/y}$, as appropriate, to restore circular symmetry before image-similarity evaluation and maximum-likelihood reconstruction.

This rescaling procedure is particularly important for higher-order subspaces, since studies on structured-light metrology have shown that the fidelity of higher-order modes tends to decrease more rapidly in the presence of spatial mismatches \cite{santos2025optimal}.

%\nocite{*}

\bibliography{bibfile}
\bibliographystyle{apsrev4-2}

\end{document}